\documentclass[aps,pre,superscriptaddress,reprint,floatfix]{revtex4-1}
\usepackage{amsmath,amssymb}
\usepackage{graphicx}% Include figure files
\usepackage{dcolumn}% Align table columns on decimal point
\usepackage{bm}% bold math
\usepackage{nameref}
\usepackage{hyperref}
\usepackage{courier}
\usepackage{braket}
\usepackage[rightcaption]{sidecap}
\usepackage{float}
\usepackage[usenames,dvipsnames]{xcolor}

\usepackage[most]{tcolorbox}
\newtcolorbox{highlighted}{colback=yellow,coltext=red,breakable}
\usepackage{lipsum}
\usepackage[normalem]{ulem}

\hypersetup{
 colorlinks   = true,	%Colours links instead of ugly boxes
  urlcolor     = blue, %Fuchsia,%Colour for external hyperlinks
  linkcolor    = blue, %Fuchsia, %Colour of internal links
  citecolor   = blue %DarkOrchid %Colour of citations#NavyBlue,Fuchsia
}

\begin{document}

\title{Long-Range Hydrodynamic Communication in Active Matter}

\author{Abhishek Thakur}
\email{abhithakur@iitb.ac.in}

\author{Bibaswan Ghosh}

\affiliation{Department of Physics, Indian Institute of Technology Bombay, Powai, Mumbai 400 076, India}

\begin{abstract}
Directed transport in active matter is typically achieved by introducing geometric asymmetry, such as ratchets or gears, that rectify the stochastic motion of active particles. Here we demonstrate a fundamentally different route in which persistent mechanical motion emerges without any geometric bias. A symmetric passive rotor placed at the center of a circular domain undergoes spontaneous unidirectional rotation driven solely by the hydrodynamic flow generated by a boundary-confined flocks of Marangoni swimmers, despite the absence of direct physical contact. While individual swimmers produce intermittent torque reversals arising from transient hydrodynamic backflow, increasing the size of the flock establishes coherent circulation that stabilizes the rotor dynamics. Particle image velocimetry (PIV) reveals that the collectively generated flow transmits momentum over macroscopic distances, converting local active forcing into robust mechanical transport. Together with a minimal flow superposition model, our results identify the suppression of reverse-torque events as a collective signature of hydrodynamic coupling between active swimmers and a passive body, providing
a contact-free mechanism for remote mechanical actuation in active matter.
\end{abstract}

\maketitle

\textit{Introduction}. Active particles comprise self-driven units that continuously convert stored or ambient energy into directed motion \cite{purcell2014life,ramaswamy2010mechanics,marchetti2013hydrodynamics,zottl2023modeling,bricard2013emergence,kumar2014flocking,bechinger2016active,boniface2019self,dey2025run,molodtsova2025micellization,thakur2026resetting}, giving rise to a broad range of nonequilibrium phenomena including collective motion \cite{bricard2013emergence,kumar2014flocking,tiwari2023capture}, spontaneous flow generation, dynamic clustering \cite{bechinger2016active,molodtsova2025micellization}, and active turbulence \cite{wensink2012meso,dunkel2013fluid,doostmohammadi2017onset,martinez2019selection}. Understanding how active suspensions transmit forces and torques to passive bodies has emerged as a central problem, not only because it provides insight into biological transport processes but also because it offers strategies for designing autonomous microfluidic devices, active engines, and flow-powered micromachines. Hydrodynamic interactions play a crucial role in these systems \cite{uchida2011generic} by mediating long-range coupling between swimmers and their surrounding environment, often producing behaviors that cannot be explained solely through direct mechanical collisions.

Among various experimental realizations of active matter, camphor-driven swimmers provide a particularly attractive macroscopic platform for studying nonequilibrium hydrodynamics \cite{bourgoin2020kolmogorovian}. Their propulsion originates from surface-tension gradients generated by asymmetric camphor dissolution. Camphor systems have revealed diverse dynamical states including oscillatory motion \cite{hayashima2001camphor,sharma2020rotational,monish2024stationary}, spontaneous rotation \cite{nakata1997self}, symmetry breaking\cite{sharma2021chimeralike}, and collective self-organization \cite{suematsu2010collective,tiwari2023capture,thakur2026beyond}, making them valuable model systems for investigating generic principles of active matter beyond microscopic biological swimmers.

Previous studies have demonstrated that active particles can rotate passive gears, transport colloids, and exert measurable stresses and torques on passive objects. In most cases, however, torque generation results primarily from repeated particle-body collisions, steric interactions, or externally imposed asymmetries \cite{di2010bacterial,chor2023many,boudet2025non}. Comparatively little is known about how purely flow-mediated interactions generated by a finite number of confined swimmers determine the statistics of torque transmission to a passive object \cite{grober2026hydrodynamic}.In particular, how the statistics of torque transmitted purely through the surrounding fluid evolve as the number of active swimmers increases has received little attention.

\begin{figure}
    \centering
    \includegraphics[width=1.0\linewidth]{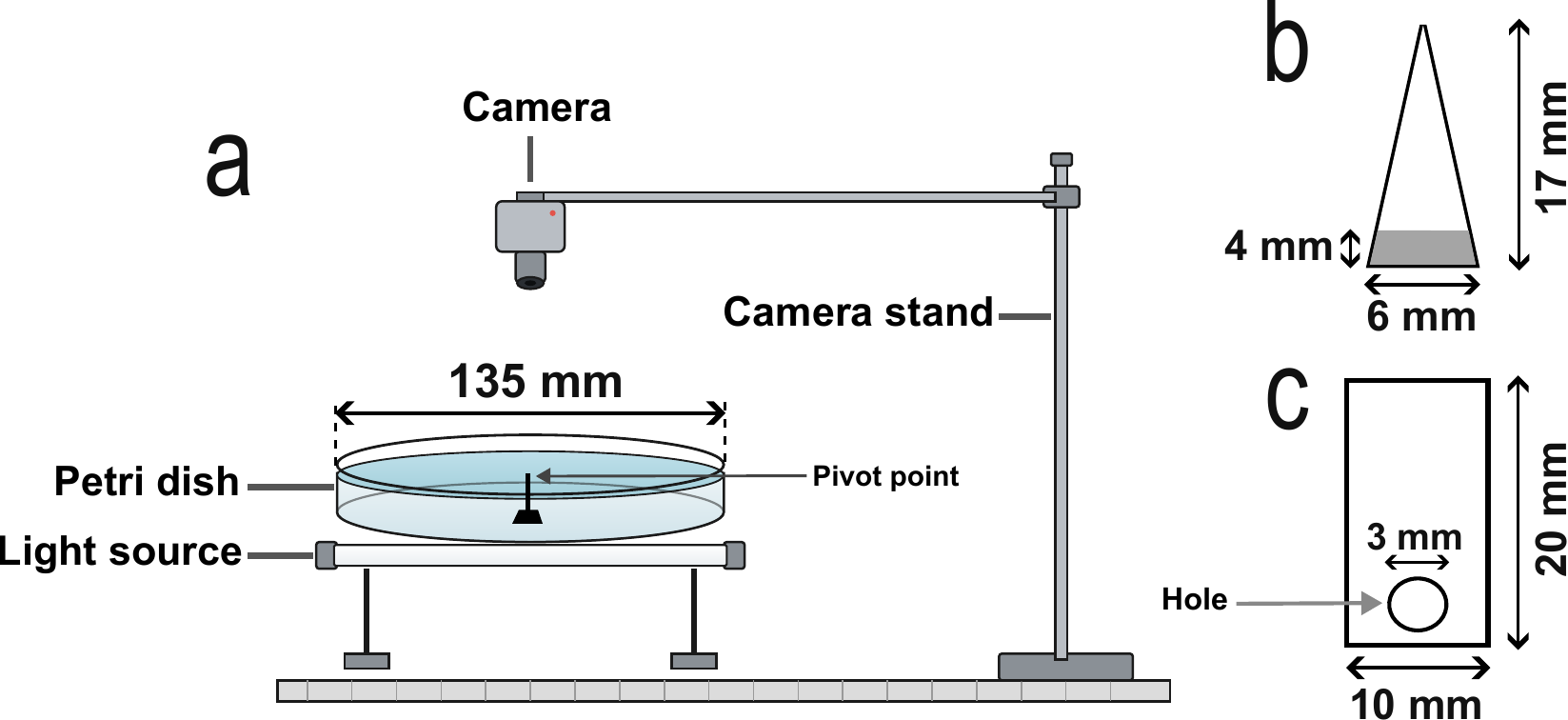}
    \caption{(a) Schematic side view of the experimental setup. (b) The active swimmer used in the setup; the grey region indicates the camphor–ethanol coated area. (c) The passive rotor.}
    \label{fig1}
\end{figure}

\begin{figure*}[t]
    \centering
    \includegraphics[width=1.0\linewidth]{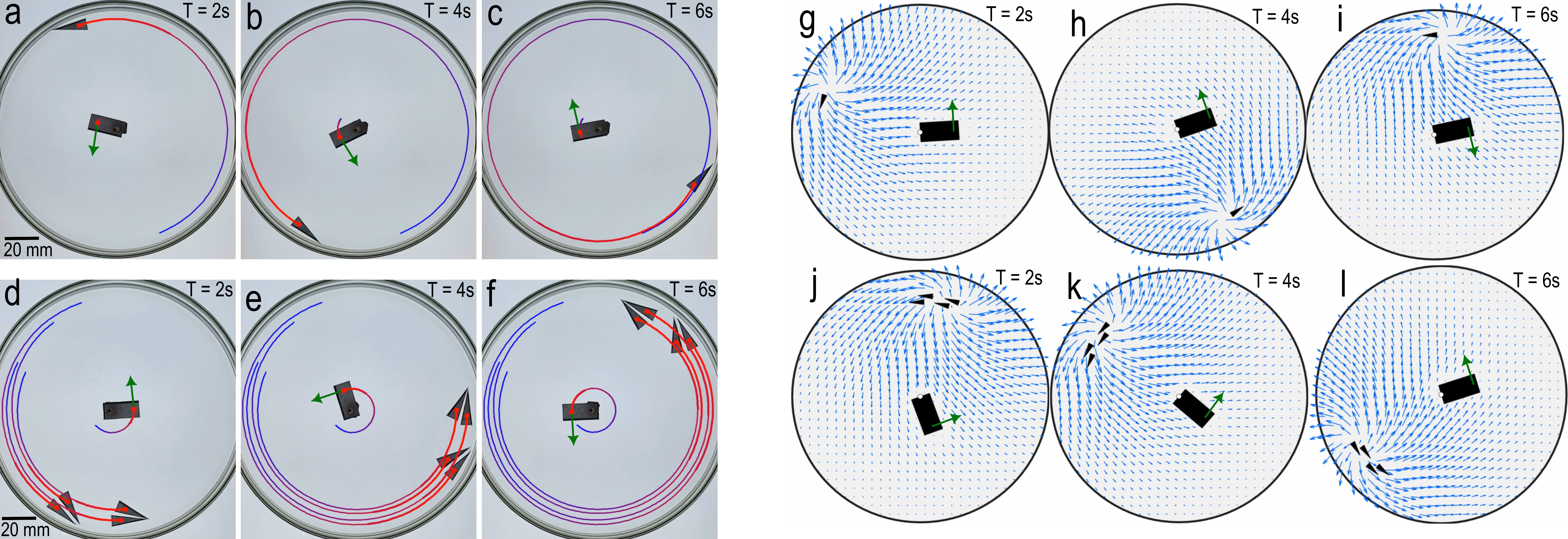}
    \caption{(a–f) Experimenta and (g–l) simulation  snapshots showing a passive rotor (rectangular stripe) pivoted at the center of a circular Petri dish and surrounded by active swimmers (triangular particles). Panels (a–c) and (g–i) show the evolution for one swimmer in the experiments and simulations, respectively, while panels (d–f) and (j–l) show the corresponding dynamics for four swimmers. Snapshots are taken at different times. Green arrows indicate the direction of motion.}
    \label{fig2}
\end{figure*}

Here we investigate this problem using self-propelled camphor swimmers (concentration 4 Molarity (4M)) confined within a circular arena containing a freely rotating passive rotor pinned at its center (Fig.~\ref{fig1}). We first examine the rotational response of the passive rotor as the number of swimmers increases. For a single swimmer, the rotor exhibits intermittent forward and reverse rotation Fig.~\ref{fig2}(a–c) (Movie1) and Fig.~\ref{fig2}(g–i) (Movie3). Notably, in Fig.~\ref{fig2}(c and i), the green arrow indicates that the rotor rotates opposite to the direction of swimmer motion, highlighting a transient reversal of the hydrodynamic torque. In contrast, for four swimmers, the rotor maintains persistent rotation in the same direction without discernible reversals over the observation period Fig.~\ref{fig2}(d–f) (Movie2) and Fig.~\ref{fig2}(j–l) (Movie4). This striking difference suggests that collective activity can stabilize the mechanical response of a passive object despite fluctuations in the underlying hydrodynamic forcing.

\begin{figure*}
    \centering
    \includegraphics[width=0.9\linewidth]{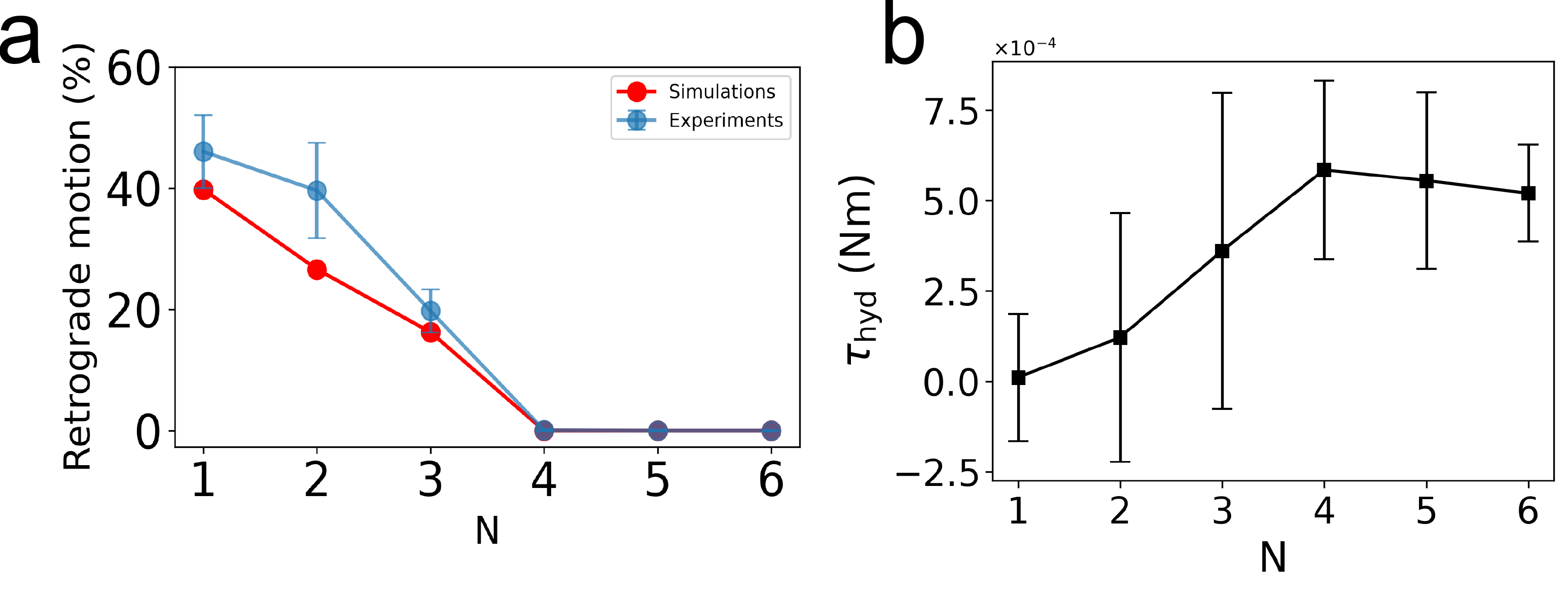}
    \caption{(a) Percentage of retrograde motion (reverse-rotation events) exhibited by the passive rotor. (b) Mean hydrodynamic torque and its fluctuation exerted on the rotor from the flow field generated by swimmers, calculated from the model.}
    \label{fig3}
\end{figure*}

The contrasting rotational dynamics observed for one and four swimmers suggest that collective hydrodynamic interactions play a crucial role in stabilizing the rotor motion. To investigate this mechanism, we develop a minimal phenomenological model describing the flow generated by multiple orbiting swimmers and its coupling to the rotational dynamics of the passive rotor.

\begin{figure*}
    \centering
    \includegraphics[width=1.0\linewidth]{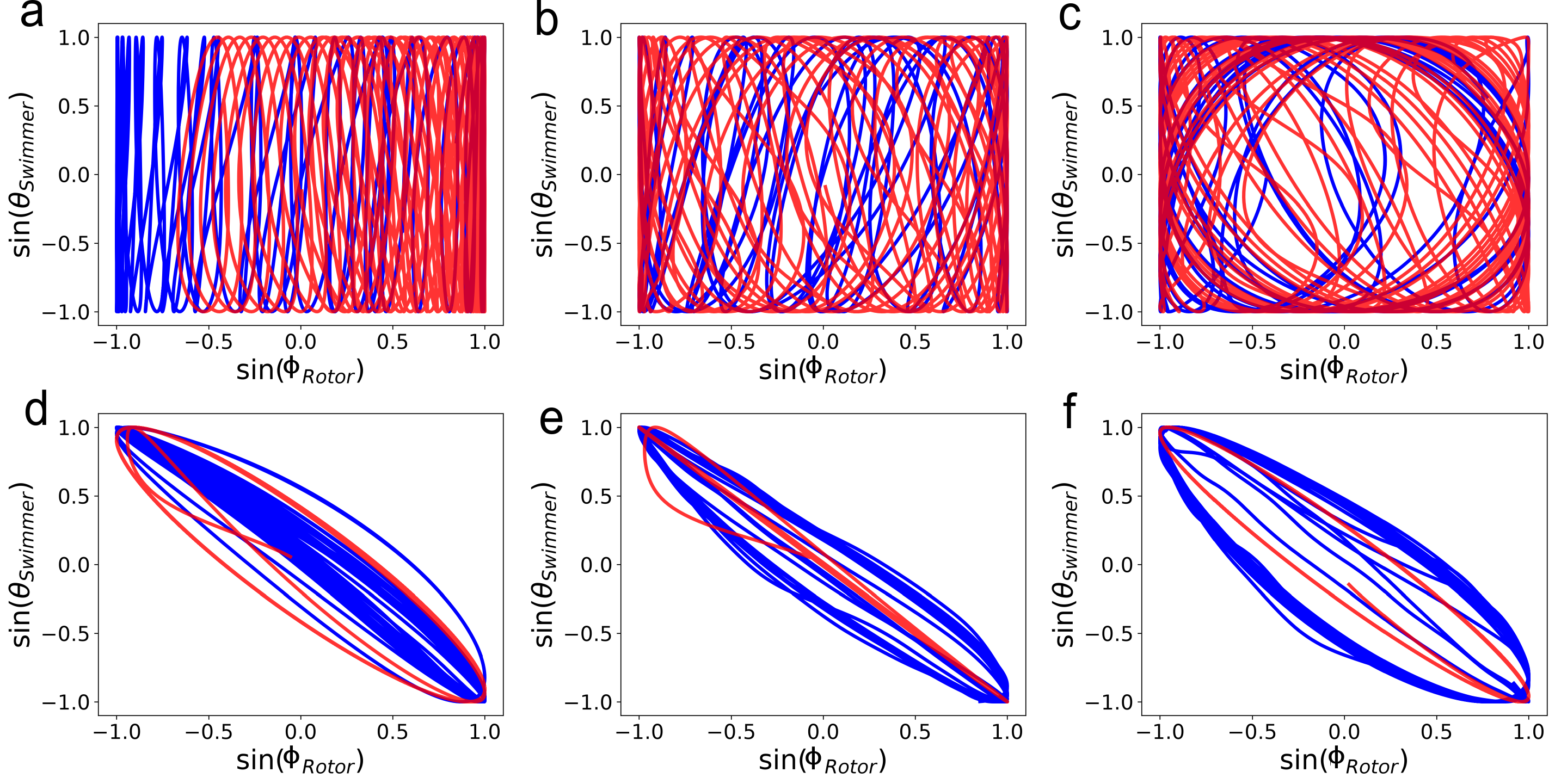}
    \caption{Phase plot for swimmers 1-6 shown in a-f respectively, red curve is from simulations and blue is of experiments, where $\theta$ ($\phi$) is the angle of the swimmer (rotor) from the center of the Petri dish ($x_{c}, y _ {c} $).}
    \label{fig4}
\end{figure*}

\textit{Model}. We consider $N$ swimmers on circular orbits of radii $R_i$ inside a dish of radius $R_\mathrm{dish}$ (taken as the unit length). Swimmer $i$ is at
$\mathbf{x}_i(t)=R_i(\cos\theta_i,\sin\theta_i)$ with angular positions prescribed as, $\theta_i(t)=\theta_i(0)+\omega t$ and $\omega>0$ corresponds to counterclockwise motion. Each swimmer drives a
localized flow centred not on the swimmer itself but at a point a distance
$\ell$ directly behind it along its trajectory,
\begin{equation}
  \mathbf{x}_{c,i}=\mathbf{x}_i+\ell\,(\sin\theta_i,\,-\cos\theta_i),
\end{equation}
which represents the camphor trail left behind the swimmer as the effective
source of the Marangoni flow. With $r_i=|\mathbf{x}-\mathbf{x}_{c,i}|$ (here, $\mathbf{x}=(x,y)$ denotes an arbitrary point in the plane of the dish,
measured from the dish centre at which the flow is evaluated). To model the localized flow generated by each swimmer, we employ a Gaussian-decaying velocity-amplitude profile, motivated by the localized vortex-core structures commonly used in vortex-flow models \cite{simpson2018detecting,shuai2022accelerated, blanco2016frequency, ducimetiere2023weakly} .The velocity magnitude associated with the (i)-th swimmer is prescribed as
\begin{equation}
  V_i(r_i)=V_{\max}\exp\!\left[-\left(\frac{r_i}{R_c}\right)^{2}\right],
\end{equation}
where $V_{\max}$ is the characteristic flow velocity and $R_c$ the decay length.
Its direction is a fixed mixture of azimuthal and radial components set by
the angle $\alpha$ \cite{nakayama2015study, bohling2010structure}. By linear superposition the total undisturbed flow is
\begin{equation}
  \mathbf{u}_0(\mathbf{x})=\sum_{i=1}^{N}V_i(r_i)\,
  \big[\cos\alpha\,\hat{\mathbf{e}}_{\theta,i}+\sin\alpha\,\hat{\mathbf{e}}_{r,i}\big],
  \label{eq:u0}
\end{equation}
where $\hat{\mathbf{e}}_{r,i}$ points away from $\mathbf{x}_{c,i}$ and
$\hat{\mathbf{e}}_{\theta,i}$ is the corresponding counterclockwise azimuthal unit vector. We use $\alpha=55^\circ$, $R_c=0.95$, $\ell=0.12$ and $V_{\max}=1$ in units of $R_\mathrm{dish}$ (See SI for other parameter values). The angular velocity $\omega$ is common to all swimmers in a given run,
but is decreased with $N$ to mimic the slower swimmers observed
experimentally as $N$ increases (Appendix A: Fig.~\ref{fig:speed} and Fig.~\ref{fig:speed_4_6}).

\begin{figure*}
    \centering
    \includegraphics[width=1.0\linewidth]{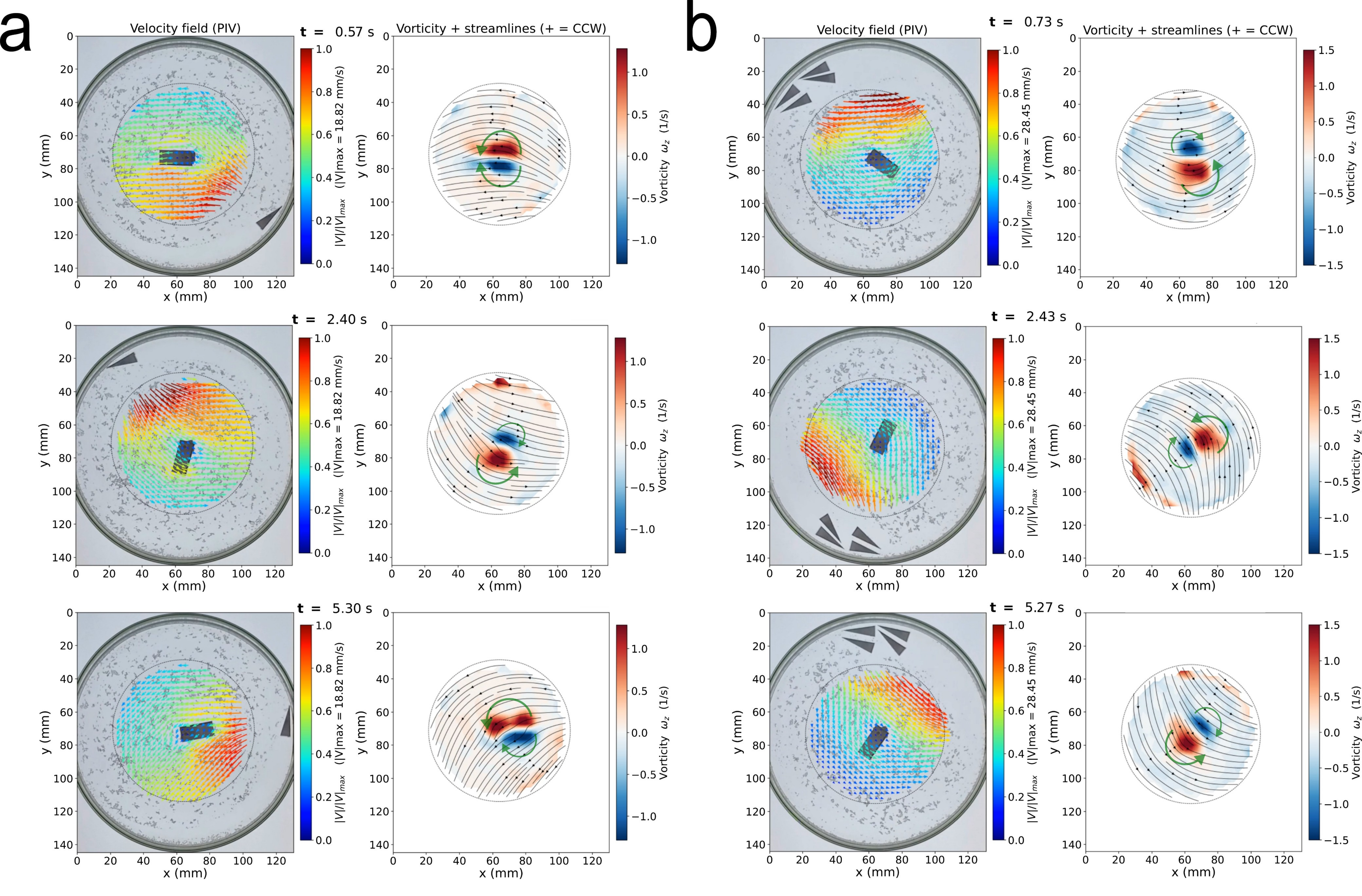}
    \caption{Experimental flow-field and vorticity measurements around the rotor using hollow glass microspheres as tracer particles. (a) Single-swimmer case, showing the evolution of the velocity field and vorticity distribution at different time instants as the swimmer crosses the rotor. (b) Four-swimmer case, showing a more persistent vorticity structure, with counterclockwise (CCW) vorticity near the rotor tip and clockwise (CW) vorticity near its lower side (shown by green arrow). The contrasting flow field structures reveals a weak hydrodynamic coupling for a single swimmer and sustained anti-phase like synchronization for four swimmers.}
    \label{fig5}
\end{figure*}

We then consider a rectangular rotor of length \(L\) and width \(W\), pivoted at the center of the arena. Its orientation is described by \(\phi(t)\), with angular velocity \(\Omega=\dot{\phi}\). A point on the rotor is parameterized by its longitudinal and transverse coordinates \((s,n)\), measured relative to the pivot. The position vector of a point on the rotor, measured from the pivot, is given by \(\mathbf r_b\). Then the local velocity of the rotor is calculated by rigid-body rotation, \(\mathbf{v}_b=\Omega\hat{\mathbf z}\times\mathbf r_b\). The relative velocity between the surrounding fluid and the rotor is therefore \(\mathbf v_{\mathrm{rel}}=\mathbf u_0-\mathbf v_b\), where \(\mathbf u_0\) is the superposed flow generated by the swimmers.

The hydrodynamic torque is calculated by integrating the transverse component of this relative velocity over the rotor surface, weighted by its lever arm \cite{gray1955propulsion, marcus2014measurements},

\[
\tau_{\mathrm{hyd}} =
c_{\mathrm{drag}}\int_0^L s
\left[
\frac{1}{W}\int_{-W/2}^{W/2}
\mathbf v_{\mathrm{rel}}\cdot\hat{\mathbf e}_{\perp}\,dn
\right]ds,
\]

where \(c_{\mathrm{drag}}\) is a phenomenological coupling coefficient and \(\hat{\mathbf e}_{\perp}\) is the transverse unit vector of the rotor. The rotational dynamics are subsequently obtained from the torque balance

\[
I\ddot{\phi}= \tau_{\mathrm{hyd}}-b_{\mathrm{pivot}}\dot{\phi},
\]

where \(I\) is the rotor moment of inertia and \(b_{\mathrm{pivot}}\) is the rotational damping coefficient. This model provides a minimal framework to investigate how the superposition of swimmer-generated flows modifies the torque acting on the passive rotor as the swimmer population increases (details of the model are in Supplementary Information (SI) section V).

We then quantify the suppression of reverse-rotation events by measuring the fraction of reverse motion (Retrograde motion) as a function of swimmer number, Fig.~\ref{fig3}(a). The experimental results show a systematic decrease in reverse motion, which eventually vanishes for four or more swimmers. Our simulations reproduce this behavior remarkably well, capturing the transition from intermittent to persistent unidirectional rotation. We further calculate the mean hydrodynamic torque ($\tau_{hyd}$) and its fluctuations from the model (see Fig.~\ref{fig3}(b)). While the mean $\tau_{hyd}$ increases with swimmer number up to four swimmers before approaching a plateau, its fluctuations continue to decrease beyond this point. This reduction in torque fluctuations suggests that the collective flow generated by an increasing number of swimmers provides a stronger and more persistent hydrodynamic forcing, thereby stabilizing the rotor's unidirectional motion.

The suppression of reverse rotation raises the question of whether the collective dynamics of the swimmers reorganize relative to the rotor. We therefore investigate the phase relationship between the swimmer and rotor (Fig.~\ref{fig4}). For one to three swimmers (Fig.~\ref{fig4}(a-c)), the phase trajectories remain broadly distributed, indicating the absence of a persistent phase relationship. Remarkably, for four to six swimmers (Fig.~\ref{fig4}(d-f)), the trajectories become concentrated along a well-defined relation corresponding to an approximately antiphase configuration. This indicates the emergence of phase locking between the collective swimmer dynamics and the rotor, coinciding with the disappearance of reverse rotation events.

To investigate the hydrodynamic origin of this synchronization, we performed additional experiments for the one-swimmer and four-swimmer case, using hollow glass microspheres \cite{maitra2025marangoni} as tracers to visualize the flow field generated by the active swimmers around the rotor. By tracking the motion of these microspheres, we aimed to characterize the vorticity distribution ($    \omega_z = \frac{\partial v}{\partial x} - \frac{\partial u}{\partial y}$) and flow structures responsible for the anti phase like synchronization and no synchronization behaviours observed in the two cases. For four swimmers, the vorticity field exhibits a remarkably persistent spatial organization at different observation times (Fig.~\ref{fig5}(b))(Movie6). A strong counterclockwise-vorticity region develops near the rotor tip, accompanied by a comparatively weaker clockwise-vorticity region near the pivoted end (green arrow). This spatial organization remains remarkably stable throughout the observation period, suggesting a persistent hydrodynamic structure associated with the synchronized state.

The corresponding flow field for a single swimmer is markedly different (Fig.~\ref{fig5}(a))(Movie5). The vorticity distribution evolves strongly with time: at \(t=0.57\) s, the counter-rotating regions are positioned side by side rather than near the rotor tip and pivot; at \(t=2.40\) s, a configuration resembling the four-swimmer state appears transiently, and at \(t=5.30\) s, the counter-rotating pattern reappears with its orientation reversed. Thus, unlike the persistent spatial organization observed for four swimmers, the single-swimmer system exhibits continuously evolving vortical structures. These observations suggest that collective activity stabilizes the hydrodynamic environment around the rotor, providing a physical picture for the emergence of phase locking and the suppression of reverse rotation.

Our results demonstrate that the statistics of torque transmission constitute a sensitive probe of collective hydrodynamic interactions in active suspensions. Rather than focusing solely on the average rotation of a passive rotor, we identify the suppression of reverse-torque events as a collective signature emerging from many-body flow-mediated interactions. These findings establish a simple macroscopic platform for quantifying collective hydrodynamic coupling and provide a new route toward understanding how active suspensions convert fluctuating microscopic forcing into robust macroscopic mechanical response. A natural future direction is to analyse the rotor's fluctuating torque and angular displacement within nonequilibrium statistical mechanics, for example
through fluctuation relations and entropy production.

\textit{Acknowledgments.}
We sincerely thank Vahid Nasirimarekani for insightful discussions related to the experiments, and Anirban Sain and
Aarsh Chotalia for their useful
suggestions and discussions.
A.T. and B.G. acknowledge the financial assistance from IITB.

\bibliographystyle{apsrev4-1}
\bibliography{ref}

\appendix
\section{Speed of swimmers and passive rotor}

\begin{figure}[h]
    \includegraphics[width=1.0\linewidth]{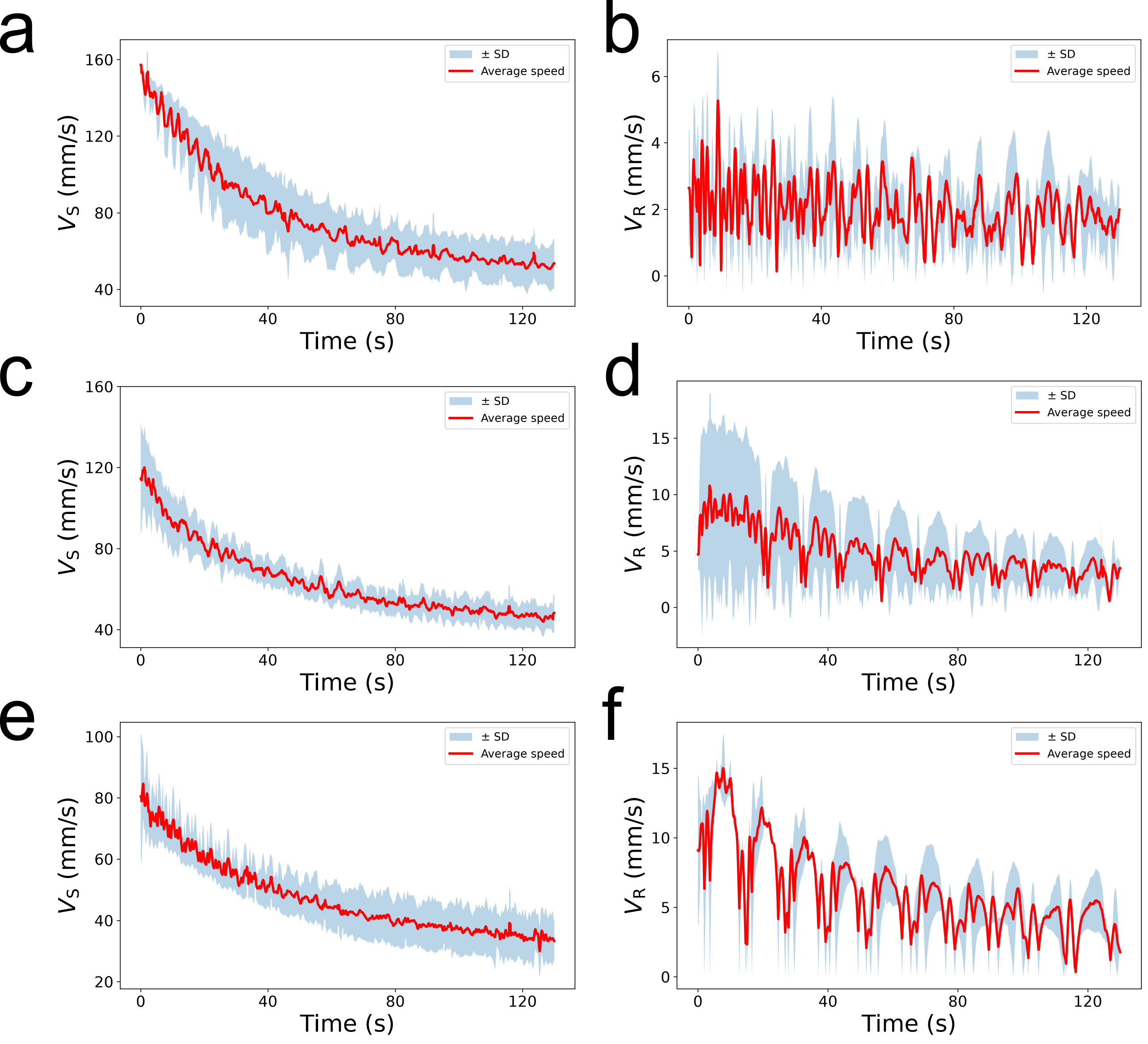}
    \caption{Panels (a), (c) and (e)  represent the mean speeds of swimmers one to three ($V_{s}$), respectively, obtained from three independent experiments at a concentration of 4M (Molarity). The red curves denote the mean speed, and the blue shaded regions represent the corresponding standard deviation. Panels (b), (d) and (f) show the corresponding speed of the passive rotor ($V_{R}$) in each case, respectively.}
    \label{fig:speed}
\end{figure}

\begin{figure}[h]
    \includegraphics[width=1.0\linewidth]{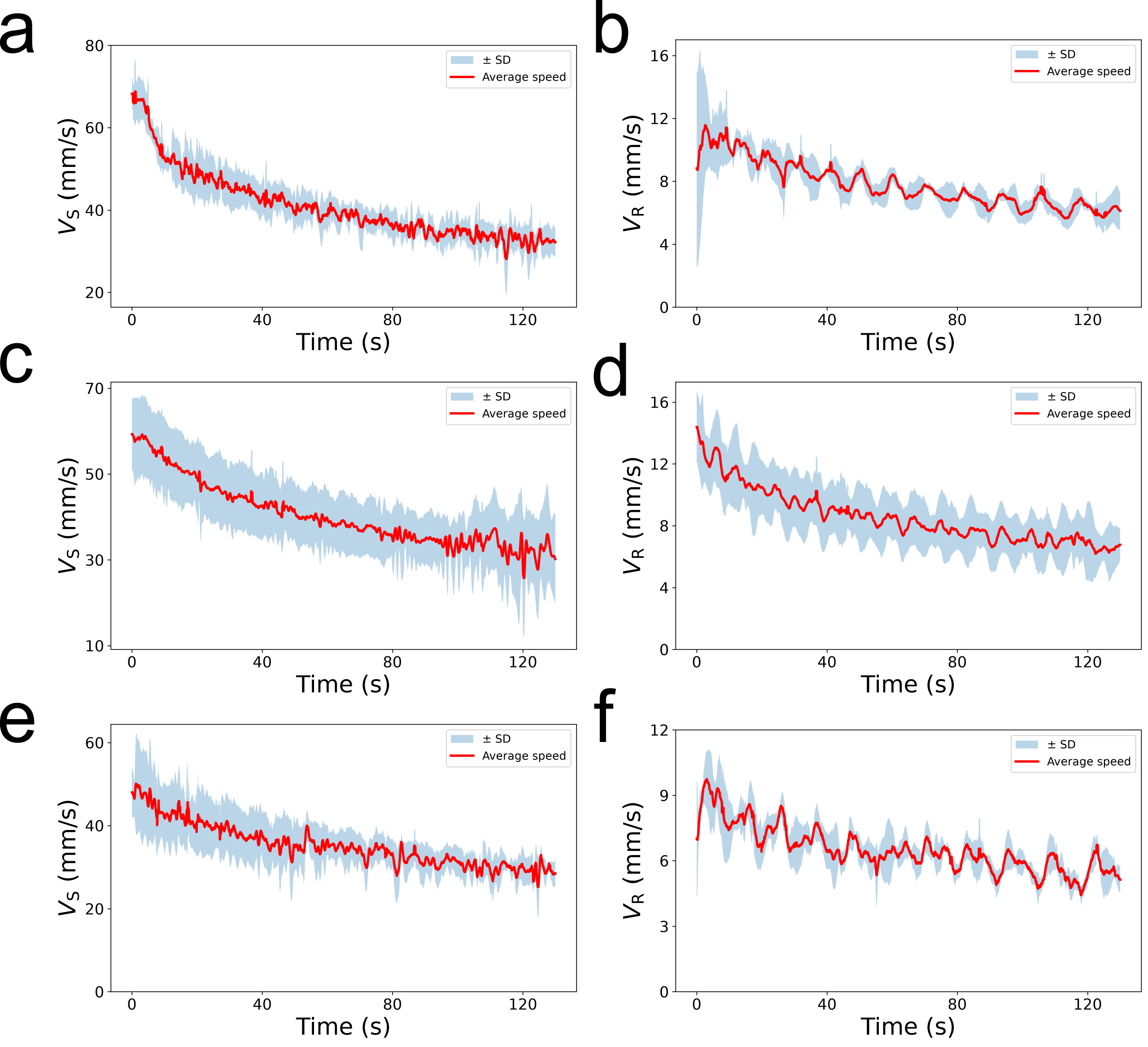}
    \caption{Panels (a), (c) and (e) represent the mean speeds of swimmers one to six ($V_{s}$), respectively, obtained from three independent experiments at a concentration of 4M (Molarity). The red curves denote the mean speed, and the blue shaded regions represent the corresponding standard deviation. Panels (b), (d) and (f) show the corresponding speed of the passive rotor ($V_{R}$) in each case, respectively.}
    \label{fig:speed_4_6}
\end{figure}

Fig.~\ref{fig:speed} and Fig.~\ref{fig:speed_4_6} shows the mean swimmer and rotor speed for three realizations along with the standard deviation for swimmer numbers, N=1-6, shown in Figs.~\ref{fig:speed}(a,c,e), and Fig.~\ref{fig:speed_4_6}(a,c,e) respectively, together with the corresponding rotor speed in Figs.~\ref{fig:speed}(b,d,f) and FIg.~\ref{fig:speed_4_6}(b,d,f). The speed of camphor-driven Marangoni swimmers is known to decrease with time due to the decay of the surface activity of camphor \cite{sharma2019rotational,upadhyaya2024stochastic}. The faster decay of the swimmer speed with increasing (N) can be attributed to the presence of multiple camphor sources\cite{gouiller2021mixing}, as the number of swimmers increases, the camphor concentration builds up more rapidly, causing the surface-tension gradient to decay faster and hence reducing the propulsion speed.

The rotor speed exhibits pronounced fluctuations for N=1, 2 and 3 (Figs.~\ref{fig:speed}(b), (f), and (j) respectively), whereas these fluctuations are significantly reduced for N=4, 5 and 6 (Figs.~\ref{fig:speed}(d), (h), and (l) respectively).

\end{document}